\documentclass[12pt]{article}
\usepackage{latexsym,amsmath,amssymb,amsbsy,graphicx}
\usepackage[space]{cite} 

\usepackage[T2A]{fontenc}
\usepackage[utf8]{inputenc}

\makeatletter\def\@biblabel#1{\hfill#1.}\makeatother

\allowdisplaybreaks
\begin {document}

\noindent\begin{minipage}{\textwidth}
\begin{center}

{\Large{\bf Kink in dual dilaton-axion theories with potential
}}\\[20pt]

{\large O.\,V. Kechkin}\\[20pt]

\parbox{.96\textwidth}{\centering\small\it
Department of General Nuclear Physics,
Faculty of Physics, M.V. Lomonosov Moscow State University, 
1(2), Leninskie gory, GSP-1, Moscow 119991, Russian Federation
\\
\ E-mail: o.v.kechkin@physics.msu.ru
}\\[1cc] 

\end{center}

{\parindent5mm The representation in terms of Ernst's complex potential is used to describe and analyze dilaton-axion theories with potential. The set of such systems is divided into pairs of dual systems with respect to the inversion of the Ernst potential. Using duality, a theory is constructed that is invariant with respect to the nonlinear Ehlers transformation. For this theory, a soliton solution is obtained that is dual to a dilaton kink in a system that is invariant with respect to the axion shift transformation.
 \vspace{6pt}\par}

\vspace{3pt}\par

\vspace{20pt}\par
\end{minipage}


\section*{Introduction}
\mbox{}\vspace{-\baselineskip}

Field-theoretic models with dilaton and axion naturally appear in superstring theory \cite{Pol}, \cite{StringsM} and have non-trivial hidden symmetry groups \cite{MS} - \cite{K}. These models include the dilaton-axion sector with the Lagrangian
\begin{equation}\label{1}
{\cal L}_0=\frac{1}{2}\left[\left(\partial\phi\right)^2+e^{2\alpha\phi}\left(\partial\kappa\right) ^2\right],
\end{equation}
where $\phi=\phi(x)$ and $\kappa=\kappa(x)$ are dilaton and axion fields, and $\alpha$ is the dilaton-axion coupling constant (for example, $\alpha=\frac {1}{2}$ for the case of low-energy heterotic string dynamics). In this paper, we consider $(1+d)$-dimensional theory on a flat background with the Minkowski metric $g_{\mu\nu}=g^{\mu\nu}=diag(1; -1, …, -1 ) $ in Cartesian orthogonal coordinates $x=\{x^{\mu}\} =\{x^0; \, x^k\}$, where $\mu=0, \dots, d$ and $k=1, \dots, d$. Here and below, for compactness, it is assumed
$\left(\partial\phi\right)^2 = g^{\mu\nu} \partial_{\mu}\phi \partial_{\nu}\phi$ (and similarly for $\left(\partial\kappa\right)^2 $).

The three-parameter group of global isotopic symmetries of the system (\ref{1}) is easily established using the Ernst potential
\begin{equation}\label{2}
F=e^{-\alpha\phi}+i\alpha\kappa,
\end{equation}
in terms of which
\begin{equation}\label{3}
{\cal L}_0=\frac{2}{\alpha^2}\frac{\partial F\partial F^*}{\left(F+F^*\right)^2}.
\end{equation}
Indeed, transformations
\begin{equation}\label{4}
F\rightarrow Fe^{-\alpha\epsilon_1}
\end{equation}
and
\begin{equation}\label{5}
F\rightarrow F+i\epsilon_2
\end{equation}
with arbitrary constant real parameters $\epsilon_1$ and
$\epsilon_2$ are explicit symmetries of (\ref{3}). Less obviously, the Lagrangian (\ref{3}) also has the discrete symmetry
\begin{equation}\label{6}
F\rightarrow F^{-1}.
\end{equation}
Finally, the remaining "hidden"\, symmetry is determined by the Ehlers transformation
\begin{equation}\label{7}
F^{-1}\rightarrow F^{-1}+i\epsilon_2;
\end{equation}
it is the result of applying the inversion (\ref{6}) to the already found continuous symmetry (\ref{5}) (and then renaming $\epsilon_2$ as $\epsilon_3$). It is also seen that under the action of (\ref{6}) the symmetry (\ref{4}) transforms into itself (with an insignificant mapping of $\epsilon_1$ to $-\epsilon_1$). Thus, the discrete symmetry (\ref{6}) splits the group of continuous symmetries (\ref{4}), (\ref{5}), (\ref{7}) into a singlet (\ref{4}) and a doublet (\ref{5}), (\ref{7}).

The Lagrangian (\ref{3}) has one more, moreover, explicit, discrete symmetry
\begin{equation}\label{8}
F\rightarrow F^*,
\end{equation}
which does not play a significant role in this article.

Note that the terminology used (Ernst potential and Ehlers transformation) is taken from works on the stationary four-dimensional General Relativity (GR) in vacuum. In this theory, a nonlinear sigma model (\ref{3}) appears with $\alpha=1$ \cite{Mazur}, \cite{ESGR}. In this case, the potential $e^{-\phi}$ plays the role of the $g_{00}$ component of the $g_{\mu\nu}$ metric of space-time, and $\kappa$ is a rotational potential. A similar correspondence between stationary vacuum GR and the electro/magnetostatic sector of Maxwell's electrodynamics with a dilaton was established in \cite{KM}.

\section{Dual dilaton-axion potentials
}
\mbox{}\vspace{-\baselineskip}

The transformation (\ref{4}) is nothing but a symmetry under the dilaton shift transformation
\begin{equation}\label{9}
\phi\rightarrow \phi+\epsilon_1,\quad \kappa\rightarrow\kappa e^{-\alpha\epsilon_1}.
\end{equation}
In the previous article \cite{DA-Sol-1} a dilaton-axion system with a potential invariant under this transformation was considered. Namely, the Lagrangian of the system was taken in the form
\begin{equation}\label{10}
L=L_0-V
\end{equation}
where the kinetic term $L_0$ is defined by the formula (\ref{1}), and an arbitrary function of an invariant combination of the fields $\phi$ and $\kappa$ can be taken as the potential $V$. As such a combination, the square of $e^{\alpha\phi}\kappa$ was taken as having an additional discrete symmetry $\kappa\rightarrow -\kappa$, which $L_0$ has (it coincides with the symmetry (\ref{8})). As a result, a class of systems (\ref{10}) was obtained with the potential $V$ as a function of the invariant $I_1=e^{2\alpha\phi}\kappa^2$ of the transformation (\ref{9}). In terms of the Ernst potential, this invariant is written as
 \begin{equation}\label{11}
I_1=-\frac{1}{\alpha^2}\left(\frac{F^*-F}{F^*+F}\right)^2;
\end{equation}
one can see that it maps to itself under the action of inversion (\ref{6}). Thus, each of the theories (\ref{10}) with the potential
\begin{equation}\label{12}
V=V\left(I_1\right)
\end{equation}
maps into itself under a discrete transformation (\ref{6}), that is, it is a singlet with respect to this transformation. 

Let us now proceed to the construction of the two remaining classes of theories of the form (\ref{10}), which are invariant under the transformations (\ref{5}) and (\ref{7}). For a theory with symmetry (\ref{5}), we put
\begin{equation}\label{13}
V=V\left(I_2\right),
\end{equation}
where 
\begin{equation}\label{14}
I_2=\frac{F+F^*}{2}
\end{equation}
can be taken as the invariant $I_2$. Finally, when constructing a class of theories that are invariant under the Ehlers transformation (\ref{7}), we put
\begin{equation}\label{15}
V=V\left(I_3\right),
\end{equation}
with
\begin{equation}\label{16}
I_3=\frac{F^{-1}+F^{*-1}}{2}.
\end{equation}
It is obvious that the invariants $I_2$ and $I_3$ transform into each other under a discrete transformation (\ref{6}). The same property is possessed (for the same functional laws $V$) by the potentials $V\left(I_2\right)$ and $V\left(I_3\right)$. As a result, we have pairs (or doublets) of systems (\ref{10}) mapped into each other under the action of inversion (\ref{6}). This circumstance allows us to speak about the discrete transformation (\ref{6}) as a duality acting on the set of dilaton-axion theories, and about dual theories related to each other by this transformation. Fundamental is the obvious fact that the duality transformation transforms the solution spaces of dual theories into each other.

Summing up, we conclude that the theory with the Lagrangian (\ref{1}), (\ref{10}), (\ref{13}), (\ref{14}) is invariant under the transformation (\ref{5}) which , as follows from the definition (\ref{2}) of the Ernst potential, is the axion shift transformation:
\begin{equation}\label{17}
\phi\rightarrow \phi,\quad \kappa\rightarrow\kappa +\epsilon_2.
\end{equation}
Then, the theory with the Lagrangian (\ref{1}), (\ref{10}), (\ref{15}), (\ref{16}) is symmetric with respect to the Ehlers transform (\ref{7}), which has nontrivial form in terms of dilaton and axion fields:
\begin{equation}\nonumber
\phi\rightarrow \phi +\frac{1}{\alpha}\log\left[1-2\epsilon_3\alpha\kappa+
\epsilon_3^2\left(e^{-2\alpha\phi}+\alpha^2\kappa^2\right)\right],
\end{equation}
\begin{equation}\label{18}
\kappa\rightarrow\frac{\kappa-\frac{\epsilon_3}{\alpha}\left(e^{-2\alpha\phi}+\alpha^2\kappa^2\right)}{1-2\epsilon_3\alpha\kappa+
\epsilon_3^2\left(e^{-2\alpha\phi}+\alpha^2\kappa^2\right)}.
\end{equation}
Finally, the duality transformation (\ref{6}), which translates the discussed classes of theories into each other, in its dilaton-axion form is a mapping 
\begin{equation}\label{19}
\phi\rightarrow \phi +\frac{1}{\alpha}\log\left(e^{-2\alpha\phi}+\alpha^2\kappa^2\right),\quad \kappa\rightarrow \frac{-\kappa}{e^{-2\alpha\phi}+\alpha^2\kappa^2}.
\end{equation}

\section{Theory with Ehlers symmetry and dualized Higgs potential
}
\mbox{}\vspace{-\baselineskip}

Let us take as the potential $V$ the function of the invariant $I$ of the following form:
\begin{equation}\label{20}
V=\frac{\lambda}{4\alpha^4}\left(\log^2I-\alpha^2v^2\right)^2,
\end{equation}
where $\lambda$ and $v$ are positive real constants. In the case of $I=I_2$ we obtain using (\ref{2}) and (\ref{14}):
\begin{equation}\label{21}
V=V_2=\frac{\lambda}{4}\left(\phi^2-v^2\right)^2,
\end{equation}
that is, the standard Higgs potential. Setting $I=I_3$, using (\ref{2}) and (\ref{16}) we obtain the potential
\begin{equation}\label{22}
V=V_3=\frac{\lambda}{4\alpha^4}\left[\log^2\left(e^{-\alpha\phi}+\alpha^2\kappa^2e^{\alpha\phi}\right)-\alpha^2v^2\right]^2.
\end{equation}
The potential (\ref{22}) can also be obtained from the Higgs potential (\ref{21}) under the action of the duality transformation (\ref{19}). By construction, the potential (\ref{22}) is symmetric with respect to the Ehlers transform (\ref{18}).

Let us now establish the vacuum structure and mass spectrum of the very non-trivial "Ehlers-Higgs dilaton-axion theory" (\ref{1}), (\ref{10}), (\ref{22}) by applying the duality transformation (\ref{19}) to the results of the corresponding analysis of the simple system (\ref{1}), (\ref{10}), (\ref{21}). Namely, it is easy to show that the constant fields
\begin{equation}\label{23}
\phi_{vac}=\pm v, \quad \kappa_{vac}=\kappa_v,
\end{equation}
where $\kappa_v$ is an arbitrary real constant, include all possible vacua of the dilaton-axion system (\ref{1}), (\ref{10}), (\ref{21}). Indeed, only on the fields (\ref{23}) is the Noetherian energy integral for this system minimized, and the equations of motion for it are identically satisfied. Then, applying the duality transformation (\ref{19}) to the field configuration (\ref{23}), we obtain a set of vacuum fields for the system (\ref{1}), (\ref{10}), (\ref{22}):
\begin{equation}\label{24}
\phi_{vac}=\pm v + \frac{1}{\alpha}\log H_v, \quad \kappa_{vac}=-\frac{\kappa_v}{H_v},
\end{equation}
Where
\begin{equation}\label{25}
H_v=e^{\mp 2\alpha v}+\alpha^2\kappa_v^2.
\end{equation}
By direct verification, one can make sure that these fields minimize the energy of the system (\ref{1}), (\ref{10}), (\ref{22}) and satisfy the Euler-Lagrange equations for it. Let us put now
\begin{equation}\label{26}
\phi=\pm v + \chi, \quad \kappa=\kappa_v+e^{\mp \alpha v}\theta
\end{equation}
for the dilaton-axion theory (\ref{1}), (\ref{10}), (\ref{21}), and calculate the quadratic part $L_2$ of its Lagrangian. The result looks like this:
\begin{equation}\label{27}
L_2=\frac{1}{2}\left(\partial\theta\right)^2+\frac{1}{2}\left(\partial\chi\right)^2 -
\frac{m^2_{\chi}}{2}\chi^2,
\end{equation}
Where
\begin{equation}\label{28}
m^2_{\chi}=2\lambda v^2.
\end{equation}
Thus, in this theory there is a massless Nambu-Goldstone mode $\theta$ and a Higgs field $\chi$ with squared mass (\ref{28}). The statement now consists in the fact that the mass spectrum of the dilaton-axion system (\ref{1}), (\ref{10}), (\ref{22}) coincides with that just found. Indeed, applying the duality transformation, we map the system (\ref{1}), (\ref{10}), (\ref{22}) into the already studied theory (\ref{1}), (\ref{10} ), (\ref{21}), after which we repeat the analysis already performed. Finally, applying duality formulas (\ref{19}) to fields (\ref{26}), and keeping only linear in $\theta$ and
$\chi$ terms results in substitutions
\begin{equation}\nonumber
\phi=\pm v + \frac{1}{\alpha}\log H_v -\frac{1}{H_v}
\left[\left(e^{\mp2\alpha v}-\alpha^2\kappa_v^2\right)\chi-2\alpha\kappa_ve^{\mp\alpha v}\theta\right],
\end{equation}
\begin{equation}\label{29}
\kappa=-\frac{\kappa_v}{H_v} -\frac{e^{\mp\alpha v}}{H_v^2}
\left[ 2\alpha\kappa_ve^{\mp\alpha v}\chi+\left(e^{\mp2\alpha v}-\alpha^2\kappa_v^2\right)\theta\right],
\end{equation}
which are guaranteed to transform the quadratic part of the Lagrangian of the system (\ref{1}), (\ref{10}), (\ref{22}) to the form (\ref{27}), (\ref{28}).

\section{Dualized dilaton kink
}
\mbox{}\vspace{-\baselineskip}

Our goal is to construct a stationary soliton in the dilaton-axion theory (\ref{1}), (\ref{10}) with a potential (\ref{22}), which we will naturally call the "Ehlers-Higgs potential". The construction process will include two stages: first, we will build a soliton in a simpler system (\ref{1}), (\ref{10}), (\ref{21}), after which we will “turn” it into a soliton in theory of interest to us using the duality transformation (\ref{19}). A soliton is understood as an exact solution of the equations of motion of the theory, which has a finite value of the energy integral.

According to Derrick's theorem, for theories of the form (\ref{1}), (\ref{10}), stationary solitons exist only for $d=1$, that is, only in the case of a two-dimensional space-time \cite{D}. We will further consider this particular case. Then $\phi=\phi(x)$, $\kappa=\kappa(x)$, where $x=x^1$ for stationary fields. The energy integral is given  by the relation
\begin{equation}\label{30}
E=\int_{-\infty}^{+\infty}dx\,\left\{\frac{1}{2}\left[\left(\phi'\right)^2+
e^{2\alpha\phi}[\left(\kappa'\right)^2\right]+V\right\},
\end{equation}
where the prime means taking the derivative with respect to $x$.

The Euler-Lagrange equations of a stationary system can also be obtained as conditions for the extremality of the functional (\ref{30}); in the case of a system with a potential (\ref{21}), they have the following form:
\begin{equation}\label{31}
\left(e^{2\alpha\phi}\kappa'\right)'=0,\quad \phi''-\alpha e^{2\alpha\phi}\left(\kappa'\right)^ 2-\lambda\left(\phi^2-v^2\right)\phi=0.
\end{equation}
First of all, we note that the fields (\ref{23}) do indeed satisfy the equations (\ref{31}) and minimize the energy of the system: the integral (\ref{30}) takes its minimum value equal to zero on them (that is, they are vacuums). Then, the conditions for the convergence of the integral (\ref{30}) at $x\rightarrow\pm\infty$ have the following form:
\begin{equation}\label{32}
\phi\rightarrow\phi_{\pm\infty},\quad \kappa\rightarrow\kappa_{\pm\infty}
\end{equation}
where $\phi_{\pm\infty}$ and $\kappa_{\pm\infty}$ are constants, and $\phi_{\pm\infty}^2=v^2$. Comparing this fact with the relations (\ref{23}), we come to the conclusion that the asymptotics of the fields must be vacua. Integrating the first of the equations (\ref{31}), we obtain: $e^{2\alpha\phi}\kappa'=C={\rm const}$, and $C=0$ due to the conditions (\ref{32}) at infinity. Thus, the axion must necessarily be constant in the case of a soliton, that is, equal to its (arbitrary) vacuum value:
\begin{equation}\label{33}
\kappa=\kappa_v
\end{equation}
(and thus $\kappa_{\pm\infty}=\kappa_v$). Substituting (\ref{33}) into the second relation in Eq. (\ref{31}), we obtain an equation for $\phi$,
\begin{equation}\label{34}
\phi''-\lambda\left(\phi^2-v^2\right)\phi=0,
\end{equation}
which is variational for the energy integral
\begin{equation}\label{35}
E=\int_{-\infty}^{+\infty}dx\,\left[\frac{1}{2}\left(\phi'\right)^2+\frac{\lambda}{4}
\left(\phi^2-v^2\right)^2\right].
\end{equation}
The soliton for such an efective system is well known: it is a kink (antikink)
\begin{equation}\label{36}
\phi=\pm v\tanh \left(\frac{x}{r_0}\right),
\end{equation}
where the sign "+" ("-") corresponds to a kink (antikink), and
\begin{equation}\label{37}
r_0=\sqrt{\frac{2}{\lambda v^2}}
\end{equation}
- "radius" of the soliton, the center of which is chosen at $x=0$. Here and below, the upper sign in the formula corresponds to the term outside the brackets, and the lower one - inside them. The mass of the kink (antikink), that is, the value of the energy integral (\ref{35}) calculated on the solution (\ref{36}), (\ref{37}), is
\begin{equation}\label{38}
E=\frac{2}{3}\sqrt{2\lambda}v^3.
\end{equation}
It is clear that the solution (\ref{33}), (\ref{36}), (\ref{37}) of the theory (\ref{1}), (\ref{10}), (\ref{21} ) has exactly the same value (\ref{38}) of the energy integral (\ref{30}).

Let us now turn to the construction of a soliton in the theory (\ref{1}), (\ref{10}), (\ref{22}), that is, in a dilaton-axion system with the Ehlers-Higgs potential. To do this, we apply the duality transformation (\ref{19}) to the constructed solution (\ref{33}), (\ref{36}), (\ref{37}). The result looks like this:
\begin{equation}\label{39}
\phi=\pm v\tanh \left(\frac{x}{r_0}\right)+\frac{1}{\alpha}\log H, \quad
\kappa=-\frac{\kappa_v}{H},
\end{equation}
where
\begin{equation}\label{40}
H=e^{\mp 2\alpha v\tanh\left(\frac{x}{r_0}\right)}+\alpha^2\kappa_v^2.
\end{equation}
The constructed fields (\ref{39}), (\ref{40}) are the exact solution of the theory (\ref{1}), (\ref{10}) with the Ehlers-Higgs potential (\ref{22}). That this is indeed a soliton solution, that is, a solution with a finite mass-energy, can be proven by referring again to the duality transformation. Namely, the density of the energy integral is the $T^0_0$-component of the energy-momentum tensor of the system. Due to the Noether procedure, which coincides with the Legendre transformation, we have $T^0_0=-L$ in the stationary case. But the Lagrangian of the theory (\ref{1}), (\ref{10}), (\ref{22}) is dual to the Lagrangian of the theory (\ref{1}), (\ref{10}), (\ref{21}), as well as the constructed solutions (\ref{39}), (\ref{40}) and (\ref{33}), (\ref{36}) in these theories. Therefore, calculating the energy integral (\ref{30}) for the solution (\ref{39}), (\ref{40}) in the theory (\ref{1}), (\ref{10}), (\ref {22}) we simply do the "inverse" duality transformation (\ref{19}), and reduce the process of this calculation to taking the integral (\ref{35}) for the solution (\ref{33}), (\ref{36}) from the theory (\ref{1}), (\ref{10}), (\ref{21}). The result is well known - it is the mass-energy of the kink (antikink) (\ref{38}). Thus, the found solution (\ref{39}), (\ref{40}) for the dilaton-axion theory (\ref{1}), (\ref{10}) with the Ehlers-Higgs potential (\ref{22}) is indeed a soliton. It is natural to call this solution "the dual image of a dilaton kink (antikink)".

Let us now dwell on some properties of the constructed soliton solution. Its asymptotics $\phi\left(\pm\infty\right)=\pm v+\frac{1}{\alpha}\log H_v,\,\, \kappa\left(\pm\infty\right)=- \frac{\kappa_v}{H_v}$, where $H_v$ is given by the formula (\ref{25}), coincide with the vacuum states (\ref{24}), (\ref{25}) of this theory. Thus, like the standard kink (antikink), its dual image “asymptotically translates” the vacuums of the dilaton-axion system into each other. In this case, the axion is a monotonic function of the $x$ coordinate, which increases (decreases) in the case of $\alpha\kappa_v<0$ ($\alpha\kappa_v>0$). As for the dilaton function, the analysis shows that it monotonically increases (decreases) when choosing
\begin{enumerate}
\item of the upper (lower) sign in formulas (\ref{39}), (\ref{40}) if the solution parameters satisfy the inequality $\left |\alpha\kappa_v\right |\geq e^{\left |\ alpha \right |v}$, and
\item of the lower (upper) sign in these formulas if $\left |\alpha\kappa_v\right |\leq e^{-\left |\alpha \right |v}$.
\end{enumerate}
In the case of $\left |\alpha\kappa_v\right |< e^{\left |\alpha \right |v}$, the dilaton function is no longer monotonic; at the point $x_0$, which is the solution of the equation
$e^{\mp \alpha v\tanh\left(\frac{x_0}{r_0}\right)}=\left |\alpha\kappa_v\right |$, it reaches its local minimum (maximum) $\phi (x_0)=
\frac{1}{\alpha}\log \left(2\left |\alpha\kappa_v\right |\right)$ for $\alpha>0$ ($\alpha<0$). The corresponding value for the axion field is $\kappa(x_0)=-\frac{1}{2\alpha^2\kappa_v}$.

It is also interesting to note that the substitution $v\rightarrow v\tanh \left(\frac{x}{r_0}\right)$ transforms the asymptotics into a solution for an arbitrary value of the coordinate both in the case of a kink (antikink) and its dual image in dilaton-axion theory with Ehlers-Higgs potential.

\section*{Conclusion}
\mbox{}\vspace{-\baselineskip}

In this paper, we have completed the program for constructing soliton solutions of the kink (antikink) type in dilaton-axion systems with a potential that does not violate one of the three independent symmetries of their common kinetic part (\ref{1}). Namely, in the previous paper \cite{DA-Sol-1} a theory with a potential with dilaton shift symmetry (\ref{4}) was investigated. Here, two systems were studied at once - with axion shift symmetry (\ref{5}) and with Ehlers symmetry (\ref{7}). The first of these two systems has a soliton in the form of a standard dilaton kink (antikink) on an arbitrary constant axion background. For the second of them, the explicit form of the potential was established, and a soliton solution dual to the dilaton kink (antikink) is obtained.

The duality transformation is a discrete transformation that is the symmetry of the kinetic part of dilaton-axion systems. It acts nontrivially on their potentials in almost all cases. However, the potential discussed in the previous article \cite{DA-Sol-1} was self-dual. It was a dilaton and axion function generalizing the Higgs potential and invariant under the duality transformation. This article also generalizes the Higgs potential, but the constructed generalizations are no longer self-dual. The duality transformation transforms the constructed potentials into each other, and together with them, the solution spaces of the corresponding dynamical systems. As a result, the soliton in a theory symmetric with respect to the Ehlers transformation was obtained as a dual image of a kink (antikink) from a simpler theory with axion shift symmetry .

The developed general formalism can be used to construct new soliton solutions. The constructed dual image of a dilaton kink (antikink) is of interest in the context of studying dilaton-axion worlds on a brane and domain walls \cite{R}, as well as the decay of a false dilaton-axion vacuum \cite{vacdec} and a number of other studies in the corresponding applications of classical and quantum field theory. Other recently constructed kink-like exact solutions can be found, for example, in \cite{kink-1}, \cite{kink-2}.


\end {document}